\documentclass[10pt]{article}

\usepackage{times}
\usepackage{amsfonts,amsbsy,eucal,mathrsfs,bbm}

\usepackage{graphicx}
\usepackage{amssymb}

\DeclareSymbolFont{lett}{U}{eur}{m}{n}
\SetSymbolFont{lett}{bold}{U}{eur}{b}{n}
\DeclareMathSymbol\LL \mathord{lett}{"03}
\DeclareMathSymbol\dd \mathord{lett}{"0E}

\newtheorem{thm}{THEOREM}
  \newtheorem{cor}[thm]{COROLLARY} 
  \newtheorem{lem}[thm]{LEMMA}
  \newtheorem{claim}[thm]{CLAIM}
  
  \newtheorem{prop}[thm]{PROPOSITION}

  \newtheorem{defn}[thm]{DEFINITION}
  \newtheorem{example}[thm]{EXAMPLE}

\newcommand{\qed}{\hfill \ensuremath{\Box}}

\begin{document}


\catcode`\@=11 

\global\newcount\nsecno \global\nsecno=0
\global\newcount\meqno \global\meqno=1
\def\newsec#1{\global\advance\nsecno by1
\eqnres@t
\section{#1}}
\def\eqnres@t{\xdef\nsecsym{\the\nsecno.}\global\meqno=1}
\def\sequentialequations{\def\eqnres@t{\bigbreak}}\xdef\nsecsym{}

\def\draftmode{\message{ DRAFTMODE } \writelabels

  {\count255=\time\divide\count255 by 60
    \xdef\hourmin{\number\count255} \multiply\count255
    by-60\advance\count255 by\time
    \xdef\hourmin{\hourmin:\ifnum\count255<10 0\fi\the\count255}}}
\def\nolabels{\def\wrlabeL##1{}\def\eqlabeL##1{}\def\reflabeL##1{}}
\def\writelabels{\def\wrlabeL##1{\leavevmode\vadjust{\rlap{\smash%
        {\line{{\escapechar=`
              \hfill\rlap{\tt\hskip.03in\string##1}}}}}}}%
  \def\eqlabeL##1{{\escapechar-1\rlap{\tt\hskip.05in\string##1}}}%
  \def\reflabeL##1{\noexpand\llap{\noexpand\sevenrm\string\string\string##1}
  }}

\nolabels

\def\eqn#1#2{ \xdef
  #1{(\nsecsym\the\meqno)}
  \global\advance\meqno by1
$$#2\eqno#1\eqlabeL#1
$$}

\def\eqalign#1{\null\,\vcenter{\openup\jot\m@th
    \ialign{\strut\hfil$\displaystyle{##}$&$\displaystyle{{}##}$\hfil
      \crcr#1\crcr}}\,}

\catcode`\@=12 
%

\def\a{\alpha}
\def\b{\beta}
\def\c{\chi}
\def\d{\delta}  \def\D{\Delta}
\def\e{\varepsilon} \def\ep{\epsilon}
\def\f{\phi}  \def\F{\Phi}
\def\g{\gamma}  \def\G{\Gamma}
\def\k{\kappa}
\def\l{\lambda}  \def\La{\Lambda}
\def\m{\mu}
\def\n{\nu}
\def\r{\rho}
\def\vr{\varrho}
\def\o{\omega}  \def\O{\Omega}
\def\p{\psi}  \def\P{\Psi}
\def\s{\sigma}  \def\S{\Sigma}
\def\th{\theta}  \def\vt{\vartheta}
\def\t{\tau}
\def\w{\varphi}
\def\x{\xi}
\def\z{\zeta}
\def\U{\Upsilon}
\def\CA{{\mathcal A}}
\def\CB{{\mathcal B}}
\def\CC{{\mathcal C}}
\def\CD{{\mathcal D}}
\def\CE{{\mathcal E}}
\def\CF{{\mathcal F}}
\def\CG{{\mathcal G}}
\def\CH{{\mathcal H}}
\def\CI{{\mathcal I}}
\def\CJ{{\mathcal J}}
\def\CK{{\mathcal K}}
\def\CL{{\mathcal L}}
\def\CM{{\mathcal M}}
\def\CN{{\mathcal N}}
\def\CO{{\mathcal O}}
\def\CP{{\mathcal P}}
\def\CQ{{\mathcal Q}}
\def\CR{{\mathcal R}}
\def\CS{{\mathcal S}}
\def\CT{{\mathcal T}}
\def\CU{{\mathcal U}}
\def\CV{{\mathcal V}}
\def\CW{{\mathcal W}}
\def\CX{{\mathcal X}}
\def\CY{{\mathcal Y}}
\def\CZ{{\mathcal Z}}
%

\def\V{\mathbb{V}}
\def\E{\mathbb{E}}
\def\R{\mathbb{R}}
\def\C{\mathbb{C}}
\def\Z{\mathbb{Z}}
\def\A{\mathbb{A}}
\def\T{\mathbb{T}}
\def\L{\mathbb{L}}
\def\D{\mathbb{D}}
\def\Q{\mathbb{Q}}


\def\mJ{\mathfrak{J}}
\def\mq{\mathfrak{q}}
\def\mQ{\mathfrak{Q}}
\def\mP{\mathfrak{P}}
\def\mp{\mathfrak{p}}
\def\mH{\mathfrak{H}}
\def\mh{\mathfrak{h}}
\def\ma{\mathfrak{a}}
\def\mA{\mathfrak{A}}
\def\mC{\mathfrak{C}}
\def\mc{\mathfrak{c}}
\def\ms{\mathfrak{s}}
\def\mS{\mathfrak{S}}
\def\mm{\mathfrak{m}}
\def\mM{\mathfrak{M}}
\def\mn{\mathfrak{n}}
\def\mN{\mathfrak{N}}
\def\mt{\mathfrak{t}}
\def\ml{\mathfrak{l}}
\def\mT{\mathfrak{T}}
\def\mL{\mathfrak{L}}
\def\mo{\mathfrak{o}}
\def\mg{\mathfrak{g}}
\def\mG{\mathfrak{G}}
\def\mf{\mathfrak{f}}
\def\mF{\mathfrak{F}}
\def\md{\mathfrak{d}}
\def\mD{\mathfrak{D}}
\def\mO{\mathfrak{O}}
\def\mk{\mathfrak{k}}
\def\mK{\mathfrak{K}}
\def\mR{\mathfrak{R}}
\def\sA{\mathscr{A}}
\def\sB{\mathscr{B}}
\def\sC{\mathscr{C}}
\def\sD{\mathscr{D}}
\def\sE{\mathscr{E}}
\def\sF{\mathscr{F}}
\def\sG{\mathscr{G}}
\def\sL{\mathscr{L}}
\def\sM{\mathscr{M}}
\def\sN{\mathscr{N}}
\def\sO{\mathscr{O}}
\def\sP{\mathscr{P}}
\def\sQ{\mathscr{Q}}
\def\sR{\mathscr{R}}
\def\sS{\mathscr{S}}
\def\sT{\mathscr{T}}
\def\sU{\mathscr{U}}
\def\sV{\mathscr{V}}
\def\sW{\mathscr{W}}
\def\sX{\mathscr{X}}
\def\sY{\mathscr{Y}}
\def\sY{\mathscr{Z}}

\def\bos#1{\boldsymbol{#1}}

\hyphenation{anom-aly anom-alies coun-ter-term coun-ter-terms
}

\def\tr{{\rm tr}} \def\Tr{{\rm Tr}}

\def\tilde{\widetilde} 
\def\hat{\widehat}
%

\def\grad#1{\,\nabla\!_{{#1}}\,}
\def\gradgrad#1#2{\,\nabla\!_{{#1}}\nabla\!_{{#2}}\,}
\def\ph{\varphi}
\def\psibar{\overline\psi}
\def\om#1#2{\omega^{#1}{}_{#2}}
\def\vev#1{\langle #1 \rangle}
\def\ha{{1\over2}}
\def\half{{\textstyle{1\over2}}} 
\def\roughly#1{\raise.3ex\hbox{$#1$\kern-.75em\lower1ex\hbox{$\sim$}}}

\def\rd{\partial}
\def\ha{{\textstyle{1\over2}}}
\def\fr#1#2{{\textstyle{#1\over#2}}}
\def\Fr#1#2{{#1\over#2}}
\def\fs#1{#1\!\!\!/\,}   
\def\Fs#1{#1\!\!\!\!/\,} 
\def\ato#1{{\buildrel #1\over\longrightarrow}}
\def\up#1#2{{\buildrel #1\over #2}}

\def\pr{\prime}
\def\ppr{{\prime\prime}}

\def\bari{\bar\imath}
\def\barj{\bar\jmath}
\def\mapr#1{\!\smash{\mathop{\longrightarrow}\limits^{#1}}\!}
\def\mapl#1{\!\smash{\mathop{\longleftarrow}\limits^{#1}}\!}
\def\mapbr{\!\smash{\mathop{\longrightarrow}\limits^{\bbs_+}}\!}
\def\mapbl{\!\smash{\mathop{\longleftarrow}\limits^{\bbs_-}}\!}
\def\mapd#1{\Big\downarrow\rlap{$\vcenter{#1}$}}
\def\mapu#1{\Big\uparrow\rlap{$\vcenter{#1}$}}
\def\maprd{\rlap{\lower.3ex\hbox{$\scriptstyle\bs_+$}}\searrow}
\def\mapld{\swarrow\!\!\!\rlap{\lower.3ex\hbox{$\scriptstyle\bs_-$}}}
\def\ne{\nearrow}
\def\se{\searrow}
\def\nw{\nwarrow}
\def\sw{\swarrow}
\def\etal{et al.}
\def\Ker{\hbox{Ker}\;}

\def\ket#1{\left|\bos{ #1}\right>}\vspace{.2in}
   \def\bra#1{\left<\bos{ #1}\right|}
\def\oket#1{\left.\bos{ #1}\right>}
\def\obra#1{\left<\bos{ #1}\right.}
\def\epv#1#2#3{\left<\bos{#1}\left|\bos{#2}\right|\bos{#3}\right>}
\def\qbvk#1#2{\bos{\left(\bos{#1},\bos{#2}\right)}}
\def\Hoch{{\tt Hoch}}
\def\rrd{\up{\rightarrow}{\rd}}
\def\lrd{\up{\leftarrow}{\rd}}


\title{\bf
Semi-Classical 
Quantum Fields Theories 
\\ 
and Frobenius Manifolds
}

\author{Jae-Suk Park\thanks{This work was supported by KOSEF Interdisciplinary Research Grant No. R01-2006-000-10638-0.}
\\
\\
Department of Mathematics
\\
Yonsei University
\\
Seoul 120-749
\\ 
Korea
}

\date{}
\maketitle
\begin{abstract}
We show that a semi-classical quantum field theory comes with
a versal family with the property that the corresponding 
partition function generates all path integrals 
and satisfies a system of $2$nd order differential equations determined by
algebras of classical observables. This versal family
gives rise to a notion of special coordinates that is  
analogous to that in string theories. 
We also show that for a large class of semi-classical theories, their
moduli space has the structure
of a Frobenius super-manifold.
\\
\end{abstract}

\newsec{Introduction}

Frobenius (formal super-)manifolds \cite{D} were obtained 
by abstracting properties of Topological Conformal Field Theory in
$2$-dimensions \cite{W1,DVV}, but the mathematical structure 
was observed much earlier
by K. Saito as  certain flat structures on moduli spaces of 
unfolding of isolated singularities \cite{S}.  
Saito's examples are physically relevant to
topological Landau-Ginzburg models.  Another large class of Frobenius
manifolds has been constructed by Barannikov and Kontsevich \cite{BK}
originally in the context of moduli space of topological
string B model with Calabi-Yau target space \cite{W2}. (For a concise
review of three constructions of Frobenius manifolds, see \cite{M}.)

The purpose of this paper is to study natural based moduli
spaces of semi-classical Quantum Field Theories (QFTs) in general.  The
tangent space to the moduli space at the base point is the space
of equivalence classes of observables of the given semi-classical
QFT.  We shall show that a large class of semi-classical QFTs has the
structure of (formal) Frobenius super-manifold on their moduli
space. This is the positive answer to a question posed in \cite{P1}.

In the Batalin-Vilkovisky (BV) scheme of Quantum Field Theory (QFT)
one constructs a BV-algebra $\left(\sC, \Delta, \cdot\right)$ with
an associated BV bracket $\left(\hbox{ },\hbox{ }\right)$, where
$\left(\sC,\cdot\right)$ is graded commutative algebra of functions on
the space $C$ of all fields and anti-fields and $\Delta$ is an odd
second-order operator, whose failure to be a derivation of the product
$\cdot$ is measured by the BV bracket \cite{BV, S1}.  A semi-classical
BV master action functional $S$ is a ghost number zero element $S\in
\sC^0$, whose restriction to the space of classical fields is the
usual classical action functional, and satisfies the 
so-called semi-classical
BV master equation:
$$
(S, S)=0, \qquad \Delta S =0.
$$
Semi-classical BV master action functional $S$ defines 
an operator $Q:=(S,\hbox{ })$, which satisfies $Q^2=0$ and is a
derivation of the product $\cdot$. The restriction of $Q$ to the space
of fields is the usual BRST operator. Then cohomology of the 
complex $(\sC, Q)$ corresponds to equivalence classes of
classical observables and is a graded commutative algebra
called {\it classical algebra of
  observables}.  A classical observable 
is a quantum observable if it belongs to $\hbox{Ker
}\Delta$.  Throughout this paper we assume that the cohomology of
the complex $(\sC, Q)$ is finite dimensional for each degree (the ghost
number).

It is natural to consider family of semi-classical QFTs defined by
family of semi-classical action functionals $\sS = S + \CO$ satisfying
the semi-classical master equation $\Delta \sS= (\sS, \sS)=0.$
Equivalently \eqn\master{ \eqalign{ \Delta \CO =0,\cr Q \CO
    +\Fr{1}{2}(\CO,\CO) =0.\cr } } An infinitesimal solution of the
above equation corresponds to a representative of $Q$-cohomology,
which is annihilated by $\Delta$, namely a classical observable which
is also a quantum observable.  The motivation for considering a family of
semi-classical action functional is to define and study a generating
functional of all path integrals of the given semi-classical field
theory.  The generating functional is 
the partition function of a versal semi-classical action
functional $\sS =S +\CO$ whose infinitesimal terms span all quantum
observables.

We shall call a QFT semi-classical if its master action functional
satisfies the semi-classical BV master equation and if every
equivalence class of classical observable has a representative
belonging to $\hbox{Ker }\Delta$.  The later requirement in the above
is natural since it is a necessary condition for a versal family of
semi-classical BV master action functionals.  A main result of this
paper is that this is also a sufficient condition to have a versal
family. We also construct a special versal master action
functional $\sS=S +\CO$ for which the corresponding 
partition function $\bos{\CZ}$ satisfies the following $2$nd order
differential equation:
$$
-\hbar \Fr{\rd^2\bos{\CZ}}{\rd t^\a \rd t^\b} +\CA_{\a\b}{}^\g \Fr{\rd\bos{\CZ}}{\rd t^\g}=0
$$
determined by the structure constants $\{\CA_{\a\b}{}^\g\}$ of the
classical algebras of observables parameterized by the moduli space.
This can be viewed as an alternative to the usual approach of studying
the quantum field theory, which compute path integrals via summing
over Feynman diagrams after suitable gauge fixing procedure.

Furthermore, we show that the tangent space to moduli space of
semi-classical QFTs has (i) a linear pencil of torsion free flat
connection and (ii) a flat invariant metric, provided that
semi-classical fields theory also comes with a BK-integral---an idea
introduced by Barannikov-Kontsevich \cite{BK}.  Combining (i) and
(ii), the moduli space of semi-classical QFTs with a BK-integral has the
structure of a (formal) Frobenius manifold.  This can be viewed as a
generalization of the construction of Barannikov-Kontsevich, which
applies to a smaller class of semi-classical QFTs associated with a
differential BV-algebra $(\sC, \Delta, Q, \cdot, (\hbox{ },\hbox{ }))$
satisfying with $\Delta Q$-property.

I am grateful to Dennis Sullivan, John Terilla and Thomas Tradler for
many helpful and inspiring conversations
This manuscript was prepared
in part during my visit to IHES in January and February 2007.  I also
thank the institute for its hospitality and an excellent research
environment. I would like to thank the referee whose careful
comments helped improve the presentation of this paper significantly.
Finally,   I would like to thank John Terilla for proofreading the final version
of this manuscript.

\newsec{Classical and Quantum Algebras}

In this section we introduce the notion of a semi-classical quantum
algebra, which is extracted from a semi-classical version of
Batalin-Vilkovisky (BV) quantization scheme of field theory.  We use
the Einstein summation convention throughout.  Fix a ground field
$\Bbbk$ of characteristic zero.  A $\Z$-graded super-commutative
associative algebra over $\Bbbk$ is a pair $(\sC, \cdot)$, where $ \sC
= \bigoplus_{k\in \Z} \sC^k $ is a $\Z$-graded $\Bbbk$-module and
$\cdot$ is a super-commutative, associative product; we say a
homogeneous element $a \in \sC^k$ carries the ghost number $k$, and
use $|a|$ to denote the ghost number of $a$. The ground field $\Bbbk$
is assigned to have the ghost number $0$.  A super-commutative and
associative product $\cdot : \sC^{k_1}\otimes
\sC^{k_2}\longrightarrow \sC^{k_1 + k_2}$ is a $\Bbbk$-bilinear map of
ghost number $0$ satisfying super-commutativity: $a\cdot b =
(-1)^{|a||b|} b\cdot a$ for any homogeneous elements $a$ and $b$, and
associativity: $a\cdot (b\cdot c)= (a\cdot b)\cdot c$.

\begin{defn}\rm
  A differential Batalin-Vilkovisky (dBV) algebra over $\Bbbk$ is a
  quadruple $(\sC, \cdot, \Delta, Q)$ where
\begin{enumerate}
\item $(\sC, \cdot)$ is a $\Z$-graded super-commutative associative
  $\Bbbk$- algebra with unit,
\item $(\sC, Q, \cdot)$ is a (super)-Commutative Differential Graded
  Algebra (CDGA) over $k$; $Q:\sC^{k} \longrightarrow \sC^{k+1}$ is a
  $\Bbbk$-linear operator of ghost number $1$, satisfying $Q^2=0$ and
  is a derivation of the product, i.e.,
$$
Q (a\cdot b)= (Q a)\cdot b +(-1)^{|a|}a\cdot (Q b),
$$
\item $\Delta:\sC^{k} \longrightarrow \sC^{k+1}$ is a $\Bbbk$-linear
  operator of ghost number $1$, satisfying $\Delta^2=0$ which is not a
  derivation of the product.  The failure of $\Delta$ to be a
  derivation of $\cdot$ is measured by the so-called BV bracket
  $(\hbox{ },\hbox{ }): \sC^{k_1}\otimes \sC^{k_2}\longrightarrow
  \sC^{k_1+k_2+1}$ defined by
$$
(-1)^{|a|}(a, b):=\Delta(a\cdot b) - \Delta a \cdot b
-(-1)^{|a|}a\cdot \Delta b,
$$
which is a derivation of the product (super-Poisson
law)
$$
(a, b\cdot c)= (a,b)\cdot c + (-1)^{(|a|+1)|b|}b\cdot (a,c),
$$
\item $\Delta Q + Q\Delta =0$.
\end{enumerate}
A BV algebra is a dBV algebra with the differential $Q=0$.
\end{defn}

\begin{prop}
For a dBV algebra $(\sC, \cdot, \Delta, Q)$ with associated 
BV bracket  $(\hbox{ },\hbox{ })$,
\begin{enumerate}
\item the triple $\left(\sC, Q, (\hbox{ },\hbox{ })\right)$ is 
a differential graded Lie algebra (DGLA) over $\Bbbk$; i.e., 
$Q:\sC^{k}\longrightarrow \sC^{k+1}$ and $Q^2=0$ (both by the Definition $1$ above)
such that 
$Q$ is a derivation the bracket
$$
Q (a, b) = (Q a, b) + (-1)^{|a|+1}(a,Q b),
$$
and satisfies the following super-commutativity
$$
(a, b) = -(-1)^{(|a|+1)(|b|+1)}(b, a),
$$
and the super-Jacobi law
$$
(a,(b,c)) = ((a,b),c) +(-1)^{(|a|+1)(|b|+1)}(b, (a,c)).
$$
\item $\Delta$ is a derivation of the BV bracket;
$$
\Delta (a, b) = (\Delta a, b) + (-1)^{|a|+1}(a,\Delta b).
$$
\end{enumerate}
\end{prop}

Proof of the above proposition is standard.  

\begin{defn}\rm
  For a dBV algebra $(\sC, \cdot, \Delta, Q)$ over $\Bbbk$ with
  associated BV bracket $(\hbox{ },\hbox{ })$, we call the quadruple
  $\bigl(\sC, \cdot, Q, (\hbox{ }, \hbox{ })\bigr)$ a classical
  algebra and we call the complex $(\sC,Q)$ a classical complex.
\end{defn}

The dBV-algebra first appeared in the context of the notion of a {\it
  semi-classical} BV master action functional as follows Consider a
BV-algebra $\left(\sC, \Delta, \cdot\right)$ with associated BV
bracket $\left(\hbox{ },\hbox{ }\right)$. A {\it semi-classical} BV
master action functional is a ghost number zero element $S\in
\sC^0$,which satisfies the semi-classical BV master equation
$(S, S)=\Delta S =0$. Defining $Q=(S,\hbox{ })$ it follows that
$\Delta Q+Q\Delta=Q^2=0$, so that the quadruple $\left(\sC,
  \cdot,\Delta, Q\right)$ is a dBV algebra with associated BV bracket
$(\hbox{ },\hbox{ })$.

Now we are ready to define the notion of a quantum algebra
in the BV quantization scheme

\begin{defn}\rm
  A quantum algebra in the BV quantization scheme (a BV quantum
  algebra) is a triple $\left(\sC[[\hbar]], \bos{K}, \cdot\right)$
  where
\begin{enumerate}
\item the pair $\left(\sC[[\hbar]], \cdot\right)$ \ is a $\Z$-graded
  super-commutative associative $\Bbbk[[\hbar]]$-algebra with unit, $\hbar$ is
  assigned to carry the ghost number $0$
\item the pair $\left(\sC[[\hbar]], \bos{K}\right)$ is a complex over
  $\Bbbk[[\hbar]]$, i.e., $\bos{K}:\sC[[\hbar]]^k\longrightarrow
  \sC[[\hbar]]^{k+1}$ and $\bos{K}^2=0$,
\item the operator $\bos{K}$ is not a derivation of the product
  $\cdot$ but the failure of $\bos{K}$ from being a derivation of the
  product is a derivation 
  divisible by $\hbar$ which is a Lie bracket, 
  i.e., for any homogeneous elements
  $\bos{a},\bos{b} \in \sC[[\hbar]]$ we have
$$
\bos{K}(\bos{a}\cdot\bos{b}) - (\bos{K} \bos{a}) \cdot \bos{b}
-(-1)^{|\bos{a}|}\bos{a}\cdot (\bos{K}\bos{a}) :=-\hbar
(-1)^{|\bos{a}|}(\bos{a},\bos{b})
$$
for some Lie bracket $\left(\hbox{ },\hbox{ }\right):
\sC[[\hbar]]^{k_1}\otimes\sC[[\hbar]]^{k_2}\longrightarrow
\sC[[\hbar]]^{k_1+k_2+1}$ satisfying $(\bos{a},\bos{b}\cdot
\bos{c})=(\bos{a},\bos{b})\cdot \bos{c}+(-1)^{(|\bos{a}+1)|\bos{b}||}
\bos{b}\cdot (\bos{a}, \bos{c})$.
\item The bracket $\left(\hbox{ },\hbox{ }\right)$ itself does not
  depend on $\hbar$, that is $\left(\sC, \sC\right) \subset \sC$.
\end{enumerate}
We call the complex $\left(\sC[[\hbar]], \bos{K}\right)$  quantum complex.
\end{defn}

It follows that the triple $\left(\sC[[\hbar]], \bos{K},(\hbox{ },
  \hbox{ })\right)$ is a DGLA over $\Bbbk[[\hbar]]$, which can be
proved as Proposition $2$.  By quantum master equation we means
$$
-\hbar\bos{K} e^{-\bos{\CO}/\hbar}=0  \Longleftrightarrow \bos{K}\bos{\CO} +\Fr{1}{2}(\bos{\CO},\bos{\CO})=0 
$$ 
for $\bos{\CO} \in \sC[[\hbar]]^0$, where we used the identity
$-\hbar\bos{K} e^{-\bos{\CO}/\hbar}=e^{\bos{\CO}/\hbar}\left(\bos{K}
  \bos{\CO} + \Fr{1}{2}(\bos{\CO},\bos{\CO})\right)$.  It also follows
that the restriction $Q = \bos{K}\bigl|_\sC$ of $\bos{K}$ on $\sC$ is
a derivation of the product.

\begin{cor}
Let $\left(\sC[[\hbar]], \bos{K}, \cdot\right)$ be a quantum algebra (over $\Bbbk[[\hbar]]$)
with the associated bracket $(\hbox{ },\hbox{ })$ and let $Q = \bos{K}\bigl|_\sC$, then 
the quadruple $(\sC, Q, \cdot, (\hbox{ },\hbox{ }))$ is a classical algebra. 
\end{cor}

\begin{prop}
Let $(\sC, \cdot, \Delta,  Q)$ be a dBV algebra over $\Bbbk$
with
associated BV bracket  $\left(\hbox{ },\hbox{ }\right)$. Then
the triple
$(\sC[[\hbar]],  \bos{K}:=-\hbar \Delta + Q, \cdot)$ is a quantum algebra.
\end{prop}

{\it Proof}. 
The condition $\Delta^2=\Delta Q +Q\Delta = Q^2=0$ implies that $\bos{K}^2=0$.
$\bos{K}$ with respect to the product satisfies the relation in Def. 4.2., as $Q$ and $\Delta$
do. We have 
$$
\bos{K}(\bos{a}\cdot\bos{b}) -\bos{K}\bos{a}\cdot \bos{b} -(-1)^{|\bos{a}|}\bos{a}\cdot(\bos{K}\bos{b})
= -\hbar (-1)^{|\bos{a}|}\left(\bos{a},\bos{b}\right)
$$
since $Q$ is a derivation of the product., etc..
\qed

A dBV algebra $(\sC, \cdot, \Delta, Q)$ with associated BV bracket
$(\hbox{ },\hbox{ })$ gives both a classical algebra $(\sC, Q,\cdot,
(\hbox{ },\hbox{ }))$ and a quantum algebra $\left(\sC[[\hbar]],
  \bos{K}=-\hbar \Delta + Q, \cdot\right)$. We call the former the
classical limit of the latter.  Cohomology of the classical complex
$(\sC, Q)$ corresponds to equivalence classes of classical
observables, while the cohomology of quantum complex $(\sC[[\hbar]],
\bos{K}=-\hbar\Delta + Q)$ corresponds to equivalence
classes of quantum observables.

\begin{defn}\rm
  A dBV algebra $(\sC, \Delta, \cdot, Q)$ is called a semi-classical
  algebra if every cohomology class of the complex $(\sC, Q)$ has a
  representative belonging to $\Ker \Delta$.
\end{defn}

For a quantum algebra formed from a semi-classical differential BV
algebra, every classical cohomology class has a representative which
is also a quantum observable.

\begin{example}\rm
  We say a dBV-algebra $(\sC, \Delta, Q, \cdot)$ has the $\Delta
  Q$-property if
$$
\left(\Ker Q \cap \Ker \Delta\right) \cap \left(\hbox{Im } \Delta \oplus \hbox{Im } Q\right) = \hbox{Im }\Delta Q
= \hbox{Im }Q\Delta.
$$
That is, for any $x\in \sC$ satisfying $Q x=0$ and $x= \Delta y$ for
some $y\in \sC$, then there exist some $z\in \sC$ such that $x =
Q\Delta z=-\Delta Q z$ etc...  Such a dBV-algebra is also
semi-classical, i.e., for any $Q$-cohomology class $\bigl[y\bigr]$
there is a representative $y^\pr$ such that $\Delta y^\pr =0$.  To see
this, consider any representative $y$ of the $Q$-cohomology class
$[y]$ such that $\Delta y=x\neq 0$; we have $Q y =0$ by definition,
which leads to $\Delta Q y=0$. Using $\Delta Q = -Q\Delta $, we have
$Q \Delta y= Q x =0$. Thus $Qx=0$ and $x=\Delta y$ and, from the
assumption there is some $z$ such that $x = \Delta y= \Delta Q z.$ It
follows that $\Delta (y - Qz)=0$ so for $y^\pr := y -Qz$, 
$\bigl[y^\pr\bigr]=\bigl[y\bigr]$ and $\Delta y^\pr=0$.
\end{example}

Note, however, that a semi-classical algebra {\it does not
  necessarily} satisfy the $\Delta Q$-property. Here is an innocent
example:

\begin{example}\rm
  Let $\sC=\Bbbk[x^1,\ldots, x^m, \eta_1,\ldots, \eta_m]$ be a
  super-polynomial algebra with free associative product subject to
  the super-commutative relations $x^i \cdot x^j= x^j\cdot x^i$,
  $x^i\cdot \eta_j=\eta_j\cdot x^i$ and $\eta_i\cdot
  \eta_j=-\eta_j\cdot\eta_i$ Assign ghost number $0$ to $\{x^i\}$ and
  $-1$ to $\{\eta_i\}$. Then $\sC = \sC^{-m}\oplus\cdots\oplus
  \sC^{-1} \oplus \sC^0$. Note that $\sC^0=\Bbbk[x^1,\ldots, x^m]$.
  Define $\Delta:= \Fr{\rd^2}{\rd x^i \eta_i}:
  \sC^{\bullet}\longrightarrow \sC^{\bullet +1}$.  It is obvious that
  $\Delta^2=0$, and the triple $(\sC, \Delta, \cdot)$ is a BV algebra
  over $\Bbbk$. We also note that $\sC^0 \in \hbox{Im }\Delta$.  To
  see this, it suffices to consider an arbitrary monomial
  $(x^1)^{N_1}\cdots (x^m)^{N_m} \in \sC^0=\Bbbk[x^1,\ldots, x^m]$ and
  observe that, for instance,
$$ (x^1)^{N_1}\cdots (x^m)^{N_m}
=\Fr{1}{(N_1+1)} \Delta\left(\eta_1\cdot (x^1)^{N_1+1}\cdots
  (x^m)^{N_m}\right).
$$

For any $S \in \sC^0$, we always have $\Delta S=(S,S)=0$. Fix $S$ and
define $Q=(S,\hbox{ })$, then  the quadruple $(\sC, \Delta, Q, \cdot)$ is a dBV algebra.
Denote $H$ the cohomology of the complex $(\sC, Q)$.
It is easy to determine that 
$$
H^0= \Bbbk[x^1,\ldots,x^m]\biggl/\left<\Fr{\rd S}{\rd x^1},\cdots, \Fr{\rd S}{\rd x^m}\right>,
$$
since $\sC^0\subset \hbox{Ker }Q$, and any element $R \in \sC^{-1}$
with the ghost number $-1$ is in the form $R =R^i \eta_i$, where
$\{R_i\}$ is a set of $m$ elements in $\sC^0$, such that $Q R = R^i
\Fr{\rd S}{\rd x^i}$.  Now we assume that $S$ is a polynomial (in
$x's$) with isolated singularities, so that the cohomology $H$ of the
complex $(\sC, Q)$ is concentrated in the ghost number zero part,
i.e., $H=H^0$. Then any representative of $H$ belongs to $\hbox{Ker
}\Delta$, since $\sC^0$ itself belongs to $\hbox{Ker }\Delta$.  Thus
the dBV algebra is obviously semi-classical.  We already know that
$\sC^0 \subset \hbox{Ker }Q \cap \hbox{Im} \Delta$, so that $\Delta
Q$-property would implies that $\sC^0 \subset \hbox{Im }Q\Delta$ and,
in particular, $H^0=0$, which is not generally true.
\end{example}

\newsec{Special Coordinates on the Moduli Space of Semi-Classical QFTs}

Consider a semi-classical algebra $(\sC,\Delta, Q, \cdot)$ with
associated BV bracket $(\hbox{ },\hbox{ })$.  Let $\{O_\a\}$ be a set
of representative of a basis of the (total) cohomology $H$ of the
complex $(\sC, Q)$ satisfying $\Delta O_\a =0$ for all $\forall \a$.
Let $\{t^\a\}$, where $t^\a \in H^*$ be the dual basis such that
$gh(t^\a) + gh(O_\a)=0$.  Denote by $\Bbbk[[H^*]]$ the algebra of
formal power series on the $\Z$-graded vector space $H^*$, the linear
dual of $H$. Then,

\begin{prop}
There exist an solution $\CO$
to the  equation
\eqn\bsh{
Q\CO +\Fr{1}{2}(\CO,\CO)=0,
}
in formal power series with values in $\sC$ 
$$
\CO = t^\a O_\a + \sum_{n=2}^\infty \Fr{1}{n!}t^{\a_n}\cdots t^{\a_1} \CO_{\a_1\cdots \a_n}
\in\left( \sC\otimes \Bbbk[[H^*]]\right)^0
$$
satisfying
\begin{enumerate}
\item {\bf Versality}: $\{O_\a\}$ forms a basis of the 
cohomology of the complex $(\sC, Q)$
\item {\bf Speciality}: for  $\CO_\a := \Fr{\rd}{\rd t^\a}\CO$ and $\CO_{\a\b} :=\Fr{\rd^2}{\rd t^\a \rd t^\b} \CO$,
$\CO_\a \in \hbox{Ker }\Delta$,
\eqn\bsha{
\eqalign{
\CO_{\a\b}&=\Delta \LL_{\a\b}\cr
\CO_\a \cdot \CO_\b &= \CA_{\a\b}{}^\g\CO_\g + Q \LL_{\a\b} + (\CO, \LL_{\a\b}).\cr
}
}
\end{enumerate}
We call such $\CO$ a special versal solution.
\end{prop}
Note that if $\CO$ is a special versal solution to equation \bsh\ then
\eqn\bshb{ \Delta \CO =0.  } Our proof is
constructive and we placed it in Section $4$ for a better
presentation.
Here we  examine some of consequences of the above proposition. We shall explain
the terminology  speciality in Section $3.2$.

\subsection{Linear Pencil of Torsion Free Flat Connections}

Begin with an obvious consequence.
\begin{cor}
  If $(\sC, \Delta, Q, \cdot,(\hbox{ },\hbox{ }))$ is a dBV algebra
  for which each cohomology class of the complex $(\sC,Q)$ has a
  representative in $\hbox{Ker }\Delta$ then the differential graded
  Lie algebra $\left(\sC, Q, (\hbox{ },\hbox{ })\right)$ is formal and
  the associated extended deformation functor (see [BK] for this
  notion) is representable by the algebra $\Bbbk[[H^*]]$.  Hence, one
  has a smooth-formal graded moduli space.
\end{cor}

Let $\CQ:= Q +\left(\CO,\hbox{ }\right): \left( \sC\otimes
  \Bbbk[[H^*]]\right)^k\longrightarrow \left( \sC\otimes
  \Bbbk[[H^*]]\right)^{k+1}$, where $\CO$ is the versal solution.
Then we have $\CQ^2=\CQ\Delta +\Delta \CQ=0$ and then the quadruple $(
\sC\otimes \Bbbk[[H^*]], \cdot, \Delta, \CQ)$ which is also a
semi-classical algebra called a versal semi-classical algebra.
Equation \bsh\ together with super-Jacobi identity implies that
$\CQ^2=0$, the super-Poisson law implies that $\CQ$ is a derivation of
the product, and the condition \bshb\ implies that $\CQ\Delta + \Delta
\CQ=0$. Also apply $\Fr{\rd}{\rd t^\a}$ to equation \bsh\ to get
$\Delta \CO_\a = \CQ\CO_\a=0$. Thus $\{\CO_\a\}$ is a set of
representatives of a basis of the total cohomology of the complex $(
\sC\otimes \Bbbk[[H^*]], \CQ)$ belonging to $\hbox{Ker }\Delta$.  The
second relation in equation \bsha\ now reads \eqn\xxx{ \CO_\a \cdot
  \CO_\b = \CA_{\a\b}{}^\g\CO_\g + \CQ \LL_{\a\b}, } thus
$\{\CA_{\a\b}{}^\g\}$ are the structure constants (in formal power
series in $\{t^\r\})$ of the algebra of the cohomology of
the versal classical complex $( \sC\otimes \Bbbk[[H^*]], \CQ)$.  It
also follows that we have corresponding versal quantum algebra
$\left(\sC[[\hbar]]\otimes \Bbbk[[\hbar]][[H^*]],
  \bos{\CK},\cdot\right)$, where $\bos{\CK}:= -\hbar\Delta
+\CQ=\bos{K} + \left(\CO,\hbox{ }\right)$ satisfying $\bos{\CK}^2=0$.

Now we state one of our main theorems, which is a corollary of Proposition
$10$.  For simplicity, we use the convention that $(-1)^{|\a|}$ stand for 
$(-1)^{|O_\a|}$.

\begin{thm}
  The set of structure constants $\{\CA_{\a\b}{}^\g\}$ in formal power
  series in $\{t^\r\}$ satisfies
$$
\eqalign{ \CA_{\a\b}{}^\g - (-1)^{|\a||\b|}\CA_{\b\a}{}^\g =0,\cr
  \CA_{\b\g}{}^\r \CA_{\a\r}{}^\s-(-1)^{|\a||\b|}\CA_{\a\g}{}^\r
  \CA_{\b\r}{}^\s=0,\cr \rd_\a \CA_{\b\g}{}^\r -(-1)^{|\a||\b|}\rd_\b
  \CA_{\a\g}{}^\r=0.  }
$$
In other words, $\left(\CA\right)_{\b}{}^\g:=dt^\a\CA_{\a\b}{}^\g$ can
be viewed as a connection $1$-form for a linear pencil of torsion-free flat
connections.
\end{thm}

{\it Proof}. The first relation in the theorem is a trivial
  consequence of the super-commutativity of the product $\cdot$;
  $\CO_\a\cdot \CO_\b = (-1)^{|\a||\b|} \CO_\b\cdot \CO_\a$.  To prove
  the second relation, consider the identity $ \CO_\a\cdot
  (\CO_\b\cdot\CO_\g) -(-1)^{|\a||\b|}\CO_\b\cdot
  (\CO_\a\cdot\CO_\g)=0 $ obtained by combining the associativity and
  super-commutativity of the product.  By applying the relation \xxx,
  and the fact that $\CQ$ is a derivation of the product, we obtain
  \eqn\xxxa{ \left(\CA_{\b\g}{}^\r
      \CA_{\a\r}{}^\s-(-1)^{|\a||\b|}\CA_{\a\g}{}^\r
      \CA_{\b\r}{}^\s\right)\CO_\s =\CQ \CM_{[\a\b]\g}, } where
$$
\eqalign{
\CM_{[\a\b]\g} :=&-\CA_{\b\g}{}^\r \LL_{\a\r} 
+(-1)^{|\a||\b|}\CA_{\a\g}{}^\r \LL_{\b\r}
\cr
&
-(-1)^{|\a|}\CO_\a\cdot \LL_{\b\g} 
+(-1)^{|\a||\b| +|\b|}\CO_\b \cdot \LL_{\a\g}.
}
$$
Thus the linear combination $ \bigl(\CA_{\b\g}{}^\r
\CA_{\a\r}{}^\s-(-1)^{|\a||\b|}\CA_{\a\g}{}^\r
\CA_{\b\r}{}^\s\bigr)\CO_\s $ vanishes in the $\CQ$-cohomology.
Furthermore every coefficient in the above linear combination must
vanish, since the $\CQ$-cohomology classes of $\{\CO_\s\}$ form a
basis of the total cohomology of the complex $( \sC\otimes
\Bbbk[[H^*]], \CQ)$.  Thus we proved the second relation in the
theorem and conclude, in turn, that $\CQ \CM_{[\a\b]\g}=0$ due to
equation \xxxa.

To prove the third relation in the theorem consider equation \xxx\ in
the form $$\CO_\b \cdot \CO_\g = \CA_{\b\g}{}^\r\CO_\r + \CQ
\LL_{\b\g}$$ and apply $\Fr{\rd}{\rd t^\a}$ to each side 
and rearrange to get \eqn\xxy{ \CC_{\a\b\g}
  = \rd_\a \CA_{\b\g}{}^\r \CO_\r + (-1)^{|\a|}\CQ\left(
    \rd_\a\LL_{\b\g}\right), } where
$$
\CC_{\a\b\g}:= \CO_{\a\b} \cdot\CO_\g +(-1)^{|\a||\b|} \CO_\b\cdot
\CO_{\a\g} -\CA_{\b\g}{}^\r \CO_{\a\r} -(\CO_\a, \LL_{\b\g}).
$$
Let $\CC_{[\a\b]\g}:= \CC_{\a\b\g} -(-1)^{|\a||\b|}\CC_{\b\a\g}$, 
which explicit expression is
$$
\eqalign{ \CC_{[\a\b]\g}:=& -\CA_{\b\g}{}^\r \CO_{\a\r} -
  \left(\CO_\a, \LL_{\b\g}\right) +(-1)^{|\a||\b|} \CO_\b\cdot
  \CO_{\a\g} \cr & +(-1)^{|\a||\b|}\CA_{\a\g}{}^\r \CO_{\b\r}
  +(-1)^{|\a||\b|} \left(\CO_\b, \LL_{\a\g}\right) - \CO_\a\cdot
  \CO_{\b\g}. 
 }
$$
Then equation \xxy\ implies that
$$
\eqalign{
\CC_{[\a\b]\g}= 
&\left(\rd_\a \CA_{\b\g}{}^\r -(-1)^{|\a||\b|}\rd_\b \CA_{\a\g}{}^\r\right)\CO_\r 
\cr
&+(-1)^{|\a|}\CQ\left(\rd_\a\LL_{\b\g}-(-1)^{|\a||\b|} \rd_\b \LL_{\a\g}\right).
}
$$
Note, the right hand side of the equation above is a $\Bbbk$-linear
combination of $\{\CO_\r\}$ plus $\CQ$-exact term.  Observe that the
third relation in the theorem would follow from looking at these
coefficients if $\CC_{[\a\b]\g} \in \hbox{Im }\CQ$.

Using $\CO_{\a\b} = \Delta \LL_{\a\b}$, the first relation in equation
\bsha\ for the {\it speciality} in Proposition $10$, one can check
that $\Delta \CM_{[\a\b]\g} = -\CC_{[\a\b]\g}$ after a direct
computation.  {}From the condition that $\CQ \CM_{[\a\b]\g}=0$, we can
express $\CM_{[\a\b]\g}$ as a $\Bbbk$-linear combination of
$\{\CO_\r\}$ plus a $\CQ$-exact term:
$$
\CM_{[\a\b]\g}= \CA_{[\a\b]\g}{}^\r \CO_\r + \CQ \LL_{[\a\b]\g},
$$
for some structure constants $\{ \CA_{[\a\b]\g}{}^\r\}$ and some
$\LL_{[\a\b]\g}$.  Apply $\Delta$ to each side above to obtain
$$
\Delta \CM_{[\a\b]\g}= \CA_{[\a\b]\g}{}^\r\Delta \CO_\r +\Delta \CQ
\LL_{[\a\b]\g} = -\CQ\Delta \LL_{[\a\b]\g},
$$
where we used $\Delta \CO_\r=0$ and $\CQ\Delta + \Delta \CQ=0$.
Combining the above with the identity $\Delta \CM_{[\a\b]\g} =
-\CC_{[\a\b]\g}$ we conclude that $\CC_{[\a\b]\g}=\CQ\Delta
\LL_{[\a\b]\g}$.  Thus $\CC_{[\a\b]\g} \in \hbox{Im }\CQ$ and the
third relation in the theorem follows.\qed

\subsection{Generating Functional for Path Integrals}

Consider the semi-classical algebra $(\sC,\cdot,\Delta, Q)$ as with a
special versal solution $\CO$ as in Proposition $10$ . Then $\CO$
is also an versal solution to the quantum master equation of the
quantum algebra $(\sC[[\hbar]], \bos{K}=-\hbar\Delta + Q,\cdot)$, i.e,
$$
-\hbar\bos{K} e^{-\CO/\hbar} =0 \Longleftrightarrow \bos{K}\CO
+\Fr{1}{2}(\CO,\CO)=0.
$$
which is equivalent to $\Delta \CO =Q\CO +\Fr{1}{2}(\CO,\CO)=0$ as
$\CO$ does not depend on $\hbar$.  Applying $\rd_\b:=\Fr{\rd}{\rd
  t^\b}$ to the above identity we have
$$
\bos{K}\left(\CO_\b\cdot e^{-\CO/\hbar}\right) =0 \Longrightarrow
\bos{K}\CO_\b + \left(\CO,\CO_\b\right)=0,
$$
which is equivalent to the conditions $\Delta \CO_\b = \CQ \CO_\b=0$.
So $\{\CO_\b\}$ is a set of representative of a basis of the total
cohomology of the complex $\left(\sC[[\hbar]]\otimes
  \Bbbk[[\hbar]][[H^*]],\bos{\CK}\right)$.  We note that the two
conditions in equation \bsha\ of Proposition $10$ imply the following
relation \eqn\bshe{ \CO_\a\cdot\CO_\b -\hbar \CO_{\a\b}=
  \CA_{\a\b}{}^\g \CO_\g +\bos{\CK} \LL_{\a\b}, } where $\bos{\CK}
\LL_{\a\b} =-\hbar\Delta \LL_{\a\b} +\CQ\LL_{\a\b}$, since, for
instance, $\Delta \LL_{\a\b}= \CO_{\a\b}$.  We also note that the
above relation may also be rewritten in the following form \eqn\bshd{
  \hbar^2 \Fr{\rd^2}{\rd t^\a \rd t^\b}\ e^{-\CO/\hbar}
  =-\hbar\CA_{\a\b}{}^\g \Fr{\rd}{\rd t^\g} e^{-\CO/\hbar} +\bos{K}
  \left(\LL_{\a\b} \cdot e^{-\CO/\hbar}\right).  }

Now we illustrate what is special about a special versal solution.
Let $\CO$ be an versal solution to equation \bsh, i.e., $Q\CO
+\Fr{1}{2}(\CO,\CO)=0$, but without the speciality condition.  Then
$\CO$ may not, in general, satisfies $\Delta \CO=0$. Assume that $\CO$
does satisfy $\Delta \CO=0$ as well, so that $-\hbar \bos{K}
e^{-\CO/\hbar}=0$.  Then $\CO_\a \in \hbox{Ker }\Delta$ and $Q\CO_\a +
(\CO, \CO_\a)=0$, so that $\bos{\CK}\CO_\a =0$, where
$\bos{\CK}=\bos{K} +(\CO,\hbox{ })$ satisfying $\bos{\CK}^2=0$.  So
$\{\CO_\b\}$ is a set of representative of a basis of the total
cohomology of the complex $\left(\sC[[\hbar]]\otimes
  \Bbbk[[\hbar]][[H^*]],\bos{\CK}\right)$.  {}From $\hbar^2\rd_\a
\rd_\b \bos{K}e^{-\CO/\hbar}=0$, we deduce that
$\bos{\CK}\left(\CO_\a\cdot\CO_\b -\hbar \CO_{\a\b}\right)=0$. Thus
$$
\CO_\a\cdot\CO_\b -\hbar \CO_{\a\b} = \bos{\CA}_{\a\b}{}^\g \CO_\g +
\bos{\CK}\bos{\LL}_{\a\b},
$$
for some formal power series $\bos{\CA}_{\a\b}{}^\g=
\CA_{\a\b}^{(0)\g}+ \hbar \CA_{\a\b}^{(1)\g} +\cdots$ in
$\Bbbk[[\hbar]][[H^*]]$ and for some
$\bos{\LL}_{\a\b}=\LL_{\a\b}^{(0)} +\hbar\LL_{\a\b}^{(1)} +\cdots \in
\left(\sC[[\hbar]]\otimes
  \Bbbk[[\hbar]][[H^*]]\right)^{|\CO_\a|+|\CO_\b|-1}$.  Then the above
relation implies that (in the classical limit) $ \CO_\a\cdot\CO_\b =
\CA_{\a\b}^{(0)\g}\CO_\g + \CQ \LL_{\a\b}^{(0)} $ so that the formal
power series $\CA_{\a\b}^{(0)\g}$ are the structure constants the
super-commutative associative $\Bbbk[[H^*]]$-algebra structure on the
cohomology of the versal classical complex $(\sC\otimes
\Bbbk[[H^*]],\CQ)$.  Finally, if $\CO$ is a special versal solution
as well, then $\CO_\a \in \hbox{Ker }\Delta$ and $\CO_{\a\b} = \Delta
\LL_{\a\b}{}^{(0)}$, and we have
$\bos{\CA}_{\a\b}{}^\g=\CA_{\a\b}^{(0)\g}$.

\def\pfi{\int \!\!\!\!\!\!\!\!\sP\;}

In the BV quantization scheme the quantum master equation is
interpreted as the condition that path integral does not depends of a
choice of gauge fixing.  More precisely, a gauge fixing corresponds to
a choice of a Lagrangian subspace in the space of all ``fields" and
``anti-fields". The path integral is supposed to be performed over
such a Lagrangian subspace and the quantum BV master equation is
supposed to be the condition that such path integral does not depends
on a smooth change of the Lagrangian subspace.

Alternatively we can view the path integral of a Quantum Field Theory
as a linear map from $\bos{K}$ cohomology of associated quantum
algebra $(\sC[[\hbar]], \bos{K},\cdot)$ to the ground ring; this
is especially concrete 
when the $\bos{K}$-cohomology is finite dimensional for each ghost
numbers, which is an assumption that we make. In the presence of a
special versal solution $\CO$ to the (semi-classical) quantum
master equation and the associated versal quantum algebra
$\left(\sC[[\hbar]]\otimes \Bbbk[[\hbar]][[H^*]],\bos{\CK}:= \bos{K}
  +({\CO},\hbox{ }),\cdot\right)$, we may consider the left cyclic
module $\bos{\CN}:= \left(\sC[[\hbar]]\otimes
  \Bbbk[[\hbar]][[H^*]]\right)\cdot e^{-\CO/\hbar}$ generated by
$e^{-\CO/\hbar}$ and define an versal Feynman path integral as a
$\Bbbk[[\hbar]][[H^*]]$ module map, which we denote by
$$
\pfi \bos{X}\cdot e^{-\CO/\hbar}
$$
for any $\bos{X} \in \left(\sC[[\hbar]]\otimes
  \Bbbk[[\hbar]][[H^*]]\right)$ with the property that \eqn\bshe{ \pfi
  \bos{K}\left(\bos{Y}\cdot e^{-\CO/\hbar}\right)\equiv
  \pfi\left(\bos{\CK Y}\right)\cdot e^{-\CO/\hbar}=0 } for any
$\bos{Y} \in \left(\sC[[\hbar]]\otimes \Bbbk[[\hbar]][[H^*]]\right)$.
We, then, define corresponding  partition function
$\bos{\CZ}$ by
$$
\bos{\CZ}:=\pfi e^{-\CO/\hbar}
$$
From equation \bshd, we have
$$
\eqalign{ \left(\hbar^2 \Fr{\rd^2}{\rd t^\a \rd t^\b}
    +\hbar\CA_{\a\b}{}^\g \Fr{\rd}{\rd t^\g}\right)\pfi e^{-\CO/\hbar}
  =\pfi \bos{K} \left(\LL_{\a\b} \cdot e^{-\CO/\hbar}\right)}
$$
and from property \bshe\ and the definition of $\bos{\CZ}$, we
have \eqn\bscf{ \eqalign{ \left(\hbar^2 \Fr{\rd^2}{\rd t^\a \rd t^\b}
      +\hbar\CA_{\a\b}{}^\g \Fr{\rd}{\rd t^\g}\right) \bos{\CZ}=0.  }}

\subsection{The Flat Metric on the Moduli Space}

In this subsection we show that the formal moduli space associated
with a semi-classical dBV algebra naturally carries the structure of a
formal Frobenius super-manifold provided the dBV algebra is
semi-classical and comes with an additional structure called BK
integral.  This result is a generalization of the main result in
\cite{BK}, which applies to the smaller class of semi-classical dBV
algebras having the $\Delta Q$-property.  We shall follow some
relevant parts in the papers \cite{BK,M} closely.  Also we do not
attempt to explain formal Frobenius super-manifolds, but refer the
reader to \cite{BK}.

\begin{defn}\rm
  Let $(\sC, \Delta, Q,\cdot)$ be a differential BV algebra over
  $\Bbbk$ with associated BV bracket $(\hbox{ },\hbox{ })$.  An even
  $\Bbbk$-linear functional $\int:\sC\longrightarrow \Bbbk$ is called
  BK integral if it satisfies the following two properties;

  (i) $\forall a, b \in \sC$ we have $\int (Q a)\cdot b = -(-1)^{|a|}
  \int a \cdot (Q b)$,

  (ii) $\forall a,b \in \sC$ we have $ \int( \Delta a)\cdot b
  =(-1)^{|a|} \int a \cdot (\Delta b)$.
\end{defn}

\begin{cor}
  Let $(\sC, \Delta, Q,\cdot)$ be a semi-classical dBV algebra over
  $\Bbbk$ with a BK integral $\int$. Assume that $\CO\in (\sC\otimes
  \Bbbk[[H^*]])^0$ is an versal solution of semi-classical master
  equation $\Delta\CO=Q\CO +\Fr{1}{2}(\CO,\CO)=0$, then $\int$ is also
  an BK integral of an versal semi-classical algebra $(\sC\otimes
  \Bbbk[[H^*]],\Delta, \CQ=Q +(\CO,\hbox{ }),\cdot)$ as an even
  $\Bbbk[[H^*]]$-linear functional $\int:\sC\otimes
  \Bbbk[[H^*]]\longrightarrow \Bbbk[[H^*]]$.
\end{cor}
{\it Proof}.
  It is suffice to show that $\int \CQ a=0$ for $\forall a \in
  \sC\otimes S(H^*)$;
$$
\int \CQ a =\int Q a + \int (\CO, a) = \int (\CO, a)=\int \Delta
(\CO\cdot a) -\int \CO\cdot \Delta a =- \int (\Delta \CO)\cdot a=0,
$$
where we used $\int Q b=\int \Delta b=0$ for any $b$ by (i) and (ii)
after setting $a=1$ and $\Delta \CO=0$ by assumption. Note also that
$\CQ$ is a derivation of the product $\cdot$, implying that $0=\int
\CQ(a\cdot b)=\int (\CQ a)\cdot b +(-1)^{|a|}\int a\cdot (\CQ b)$.
\qed

Consider a semi-classical dBV algebra with a BK integral $\int$.  Let
$\CO$ be a special versal solution as in Proposition $10$, so that
$\CO_\a :=\rd_\a\CO \in \hbox{Ker }\Delta$ and $\CO_{\a\b}=\rd_\a
\CO_\b = \Delta \LL_{\a\b}$.
 
\begin{claim}
  Denote $g_{\a\b}:= \int \CO_\a\cdot\CO_b$ and $\CA_{\a\b\g} :=
  \CA_{\a\b}{}^\r g_{\r\g}$.  Then $\rd_\g g_{\a\b}=0$ and
  $\CA_{\a\b\g}= (-1)^{|\a|(|\b|+|\g|)}\CA_{\b\g\a}$.
\end{claim}

{\it Proof}. 
  Applying $\rd_\g:= \Fr{\rd}{\rd t^\g}$ to the definition of
  $g_{\a\b}$, yields
$$
\eqalign{ \rd_\g g_{\a\b}&= \int (\rd_\g\CO_\a)\cdot\CO_b +
  (-1)^{|\a||\g|}\int \CO_\a\cdot(\rd_\g\CO_b) \cr &= \int (\Delta
  \LL_{\g\a})\cdot\CO_\b + (-1)^{|\a||\g|}\int
  \CO_\a\cdot(\Delta\LL_{\g\b}) \cr &=0, }
$$
since $\Delta \CO_\a=0$ for all $\a$ and $\int (\hbox{Im }
\Delta)\cdot(\hbox{Ker }\Delta)=0$.  For the second claim we first
note that
$$
\int (\CO_\a\cdot \CO_{\b})\cdot \CO_\g=\int (\CA_{\a\b}{}^\r \CO_\r
+\CQ\LL_{\a\b}) \cdot \CO_\g = \CA_{\a\b}{}^\r\int \CO_\r\cdot \CO_\g
= \CA_{\a\b}{}^\r g_{\r\g}=\CA_{\a\b\g},
$$
where we used $\CQ\CO_\g =0$ and the fact that $\CQ$ is a derivation
of the product.  Similarly,
$$
(-1)^{|\a|(|\b|+|\g|)} \int (\CO_{\b}\cdot \CO_{\g})\cdot \CO_\a
=(-1)^{|\a|(|\b|+|\g|)}\CA_{\b\g\a}.
$$
So the second claim is equivalent to
$$
\int(\CO_\a\cdot \CO_{\b})\cdot \CO_\g = (-1)^{|\a|(|\b|+|\g|)}
\int(\CO_{\b}\cdot \CO_{\g})\cdot \CO_\a,
$$
which is obvious due the super-commutativity and associativity of the
product.\qed

Combining the identities in Theorem $12$ and the above Claim, we have
\begin{thm}
  The moduli space of a semi-classical QFT equipped with a BK integral
  carries a structure of formal Frobenius super-manifold.
\end{thm}

It seems natural to expect that there be a mirror phenomenon of
semi-classical QFT with BK integral involved in the general study of
morphisms of formal Frobenius structures.

\newsec{Proof of Proposition 10} We begin by explaining our plan for
proof.  Instead of solving the semi-classical master equation \bsh\
directly we concurrently build a pair of inductive systems
$$ 
\matrix{ I_0 &\subset& I_1&\subset& I_2&\subset& \cdots& \subset&
  I_{n-1}&\subset& I_{n}&\subset& \cdots \cr \downarrow&\nearrow&
  \downarrow&\nearrow& \downarrow&\nearrow& \cdots& \nearrow&
  \downarrow&\nearrow& \downarrow&\nearrow& \cdots \cr J_0 &\subset&
  J_1&\subset& J_2&\subset& \cdots& \subset& J_{n-1}&\subset&
  J_{n}&\subset& \cdots \cr }
$$

The inductive systems consist of
\begin{itemize}
\item $I_0=\left(\left\{\CO^{[0]}_\a\right\}\right)$: fix a set
  $\{O_\a\}$ of representative of a homogeneous basis of the
  cohomology of the complex $(\sC, Q)$ satisfying $\Delta O_\a =0$ and
  define $\CO^{[0]}_\a: =O_\a$.

\item
  $J_{n-1}=\left(\left\{\LL^{[0]}_{\a\b},\CA^{[0]}_{\a\b}{}^\g\right\},
    \left\{\LL^{[1]}_{\a\b},\CA^{[1]}_{\a\b}{}^\g\right\} \ldots,
    \left\{\LL^{[n-1]}_{\a\b},\CA^{[n-1]}_{\a\b}{}^\g\right\}\right)$,
  where $\CA_{\a\b}^{[k]\g}$ is the part of the algebra structure
  constant of word-length $k$ in $\{t^\r\}$ and $\LL^{[k]}_{\a\b} \in
  \left(\sC\otimes S^k(H^*)\right)^{|O_\a|+|O_\b| -1}$, which together
  satisfy the following system of equations \eqn\asumb{ Q
    \LL^{[\ell]}_{\a\b}= \sum_{j=0}^{\ell} \CO^{[j]}_\a \cdot
    \CO^{[\ell-j]}_\b - \sum_{j=1}^{\ell-1}\left(\CO^{[\ell-j]},
      \LL^{[j]}_{\a\b}\right) -
    \sum_{j=0}^{\ell}\CA_{\a\b}^{[\ell-j]\g}\CO^{[j]}_\g, } for all
  $\ell=0,1,\ldots, n-1$, where $\CO^{[\ell+1]}_\a :=
  \Fr{1}{(\ell+1)}t^\b \Delta\LL^{[\ell]}_{\b\a}$.

\item $I_n=\left(\left\{\CO^{[0]}_\a\right\},
    \left\{\CO^{[1]}_\a\right\},\cdots,
    \left\{\CO^{[n]}_\a\right\}\right)$, where $\CO^{[k]}_\a \in
  \bigl(\sC\otimes S^k(H^*)\bigr)^{|O_\a|}$, which satisfy the
  following set of equations \eqn\asuma{ Q\CO^{[k]}_\a
    =-\sum_{j=1}^{k}\left(\CO^{[j]},\CO^{[k-j]}_\a\right), } for all
  $k=0,1,\ldots,n$, where $\CO^{[k+1]} :=\Fr{1}{(k+1)} t^\a
  \CO^{[k]}_\a$
\end{itemize}

The following elementary lemma says that building $I_n$ is equivalent
to solving the Maurer-Cartan (MC) equation $Q\CO +
\Fr{1}{2}(\CO,\CO)=0$ modulo $t^{n+2}$.
\begin{lem}
  Let $\CO^{[j+1]} :=\Fr{1}{(j+1)} t^\a \CO^{[j]}_\a$ for
  $j=0,1,\ldots, n$. Let $k$ be any integer with $0\leq k \leq n$,
  then the conditions in \asuma\
  imply that
$$
Q\CO^{[k+1]} +\Fr{1}{2}\sum_{j=1}^{k}(\CO^{[k-j+1]},\CO^{[j]})=0.
$$
\end{lem}

{\it Proof}.
  Multiply $(-1)^{|\a|} t^\a$ to the both hand-sides of the given
  condition and sum over $\a$ to get
$$
\sum_{j=0}^{k-1}\left(\CO^{[k-j]},t^\a\CO^{[j]}_\a\right)=- Q
t^\a\CO^{[k]}_\a,
$$
which gives
$$
\sum_{j=0}^{k-1}(j+1)\left(\CO^{[k-j]},\CO^{[j+1]}\right)=- (k+1)Q
\CO^{[k+1]}.
$$
The left hand side of the above equation can be rewritten as
$$
\sum_{j=1}^{k}j(\CO^{[k+1-j]},\CO^{[j]})\equiv
\Fr{1}{2}(k+1)\sum_{j=1}^{k}(\CO^{[k+1-j]},\CO^{[j]}),
$$
using $(\CO^{[k+1-j]},\CO^{[j]})=(\CO^{[j]},\CO^{[k+1-j]})$ and
re-summing.\qed

We remark that the inductive system $I_0\subset I_1\subset
\cdots\subset I_n\subset$ by itself is rather standard. Assume that
one has built $I_n$, equivalently, that one has constructed a solution
of the MC equation of a DGLA modulo $t^{n+2}$. The obstruction for the
inclusion $I_n \subset I_{n+1}$ lies on the cohomology of the DGLA,
and the obstruction does not depends on the choice of particular
solution modulo $t^{n+2}$.  If the obstruction vanishes, one can build
an $I_{n+1}$ by making certain choice.
In our case the existence of a $J_{n}$ shall be used to show that (i)
the obstruction for an inclusion $I_n \subset I_{n+1}$ vanishes and
(ii) there is a special choice for $I_{n+1}$.  Specifically, $I_{n+1}$
shall be obtained by $J_{n}$ by applying $\Delta$.

Now we are going to build the above mentioned pair of inductive
systems.

\begin{itemize}

\item $I_0\longrightarrow J_0$; Let $\CL^{[0]}_{\a\b}:=\CO^{[0]}_\a
  \cdot \CO^{[0]}_\b$, defined terms of $I_0$, which is $Q$-closed
  since $Q$ is a derivation of the product. Thus we can express
  $\CL^{[0]}_{\a\b}$ as
$$
\CL^{[0]}_{\a\b}= \CA_{\a\b}^{[0]\g} \CO^{[0]}_\g + Q \LL^{[0]}_{\a\b}
$$
for some structure constants $\CA_{\a\b}^{[0]\g}$ and for some
$\LL^{[0]}_{\a\b}\in \sC^{|O_\a| +|O_\b|-1}$, which is defined modulo
$\hbox{Ker } Q$. Fix a set $\{\LL^{[0]}_{\a\b}\}$ once and for all.
Then we have $J_0=
\left(\left\{\CA^{[0]}_{\a\b}{}^\g,\LL^{[0]}_{\a\b}\right\}\right)$
satisfying \eqn\bvb{ \CO^{[0]}_\a \cdot
  \CO^{[0]}_\b-\CA^{[0]}_{\a\b}{}^\g \CO^{[0]}_\g =Q \LL^{[0]}_{\a\b},
} which is the $n=0$ case of \asumb.

\item $J_0 \rightarrow I_1$; Apply $\Delta$ to the both hand-sides
  equation \bvb\ and $\Delta O_\a =\Delta Q+ Q\Delta=0$ to get $
  (-1)^{|\a|} \left(\CO^{[0]}_\a,\CO^{[0]}_\b\right) = - Q \Delta
  \LL^{[0]}_{\a\b}$.  After multiplying $(-1)^{|\a|} t^\a$ to the both
  hand-sides of the above equation and summing over $\a$ we get
$$
\left(t^\a\CO^{[0]}_\a,\CO^{[0]}_\b\right) =
-Q\left(t^\a\Delta\LL^{[0]}_{\a\b}\right),
$$
which leads to $ (\CO^{[1]},\CO^{[0]}_\a)
=-Q\left(t^\a\Delta\LL^{[0]}_{\a\b}\right), $ where $\CO^{[1]}:=
t^\a\CO^{[0]}_\a$.  By setting $\CO^{[1]}_\b =
t^\a\Delta\LL^{[0]}_{\a\b}$ we build
$I_1=\left(\left\{\CO^{[0]}_\a\right\},\left\{\CO^{[1]}_\a\right\}\right)$
which satisfy the following system of equations: \eqn\bvc{ \eqalign{
    Q\CO^{[0]}_\a&=0,\cr Q\CO^{[1]}_\a&=-(\CO^{[1]},\CO^{[0]}_\a),\cr
  } } which is the $n=1$ case of \asuma.  Define $\CO^{[2]}=
\Fr{1}{2}t^\a \CO^{[1]}_{\a}$.  The inclusion $I_0\subset I_1$ is
obvious.
\end{itemize}

Fix $n>0$ and assume that we have $J_{n-1}$ and $I_n$ as described
before;
$$ 
\matrix{ I_0 &\subset& I_1&\subset& I_2&\subset& \cdots& \subset&
  I_{n-1}&\subset& I_{n}&&
  \cr \downarrow&\nearrow&
  \downarrow&\nearrow& \downarrow&\nearrow& \cdots& \nearrow&
  \downarrow&\nearrow&&& 
  \cr J_0 &\subset&
  J_1&\subset& J_2&\subset& \cdots& \subset& J_{n-1}&& &&
  \cr }.
$$
We shall establish
$$ 
\matrix{ I_0 &\subset& I_1&\subset& I_2&\subset& \cdots& \subset&
  I_{n-1}&\subset& I_{n}&\subset&I_{n+1}\cr \downarrow&\nearrow&
  \downarrow&\nearrow& \downarrow&\nearrow& \cdots& \nearrow&
  \downarrow&\nearrow& \downarrow&\nearrow& \cr J_0 &\subset&
  J_1&\subset& J_2&\subset& \cdots& \subset& J_{n-1}&\subset& J_{n}&&
  \cr }.
$$
\begin{itemize}
\item $I_{n}\longrightarrow J_n$; Let $\CL^{[n]}_{\a\b}$ be given by
$$
\CL^{[n]}_{\a\b}:= \sum_{k=0}^{n} \CO^{[k]}_\a \cdot \CO^{[n-k]}_\b -
\sum_{k=0}^{n-1}\left(\CO^{[n-k]}, \LL^{[k]}_{\a\b}\right) -
\sum_{k=1}^{n}\CA_{\a\b}^{[n-k]\g}\CO^{[k]}_\g,
$$
which is a combination of $I_n$ and $J_{n-1}$.  Now we show that $Q
\CL^{[n]}=0$.  Using the fact that $Q$ is a derivation of both the
product and the bracket, we have
$$
\eqalign{ Q\CL^{[n]}_{\a\b}:= & \sum_{j=0}^{n} Q\CO^{[j]}_\a \cdot
  \CO^{[n-j]}_\b +(-1)^{|\a|}\sum_{j=0}^{n} \CO^{[n-j]}_\a \cdot
  Q\CO^{[j]}_\b \cr &- \sum_{j=0}^{n-1}\left(Q\CO^{[n-j]},
    \LL^{[j]}_{\a\b}\right) + \sum_{j=0}^{n-1}\left(\CO^{[n-j]},
    Q\LL^{[j]}_{\a\b}\right) - \sum_{j=1}^{n}\CA_{\a\b}^{[n-j]\g}
  Q\CO^{[k]}_\g. }
$$
Now we use assumptions \asuma\ and \asumb\ to obtain
$$
\eqalign{ Q\CL^{[n]}_\a=& \sum_{k=0}^{n-1}\sum_{j=0}^{k}\biggl(
  \left(\CO^{[n-k]},\CO^{[j]}_\a \cdot \CO^{[k-j]}_\b \right) -
  \CO^{[j]}_\a \cdot\left(\CO^{[n-k]}, \CO^{[k-j]}_\b \right) -
  \left(\CO^{[n-k]},\CO^{[j]}_\a \right) \cdot \CO^{[k-j]}_\b \biggr)
  \cr & - \sum_{k=1}^{n}\left(Q\CO^{[k]}+ \Fr{1}{2} \sum_{j=1}^{k-1}
    \left(\CO^{[j]}, \CO^{[k-j]}\right), \LL^{[n-k]}_{\a\b}\right), }
$$
after some simple re-summations, an application of the super-Jacobi
identity and one cancellation between two terms involving
$\CA_{\a\b}^{[j]\g}$.  The first line of the right-hand sides of the
above vanishes due to the super-Poisson law, and the second line also
vanishes due to the identity \asuma. Thus $Q\CL^{[n]}_{\a\b}=0$.

Consequently we can express $\CL^{[n]}_{\a\b}$ in terms of the set of
fixed representatives $\{\CO^{[0]}_\g\}$ of the cohomology classes of
the complex $(\sC,Q)$ modulo $Q$-exact terms:
$$
\CL^{[n]}_{\a\b}= \CA_{\a\b}^{[n]\g}\CO^{[0]}_\g +Q \LL^{[n]}_{\a\b},
$$
where $\left\{\CA_{\a\b}^{[n]\g}\right\}$ is a set of structure
constants with word-length $n$ and $\LL^{[n]}_{\a\b} \in
\left(\sC\otimes S^n(H^*)\right)^{|O_\a|+|O_\b| -1}$ is defined modulo
$\hbox{Ker } Q$. We fix a set $\{\LL^{[n]}_{\a\b}\}$ once and for all.
Now we set $$J_n:=\left(J_{n-1},
  \left\{\CA_{\a\b}^{[n]\g},\LL^{[n]}_{\a\b}\right\}\right)$$ 
which satisfy, for all $k=0,1,\ldots, n$ \eqn\enik{ \sum_{j=0}^{k}
  \CO^{[j]}_\a \cdot \CO^{[k-j]}_\b -
  \sum_{j=0}^{k-1}\left(\CO^{[k-j]}, \LL^{[j]}_{\a\b}\right) -
  \sum_{j=1}^{k}\CA_{\a\b}^{[k-j]\g}\CO^{[j]}_\g
  =\CA_{\a\b}^{[k]\g}\CO^{[0]}_\g +Q \LL^{[k]}_{\a\b} } by definition.
The inclusion $J_{n-1}\subset J_n$ is obvious.

\item $J_{n}\rightarrow I_{n+1}$ amounts to setting $ \CO^{[n+1]}_\a =
  \Fr{1}{(n+1)}t^\b \Delta\LL^{[n]}_{\b\a} $ such that
$$
\eqalign{ I_{n+1}:=\left(I_{n}, \left\{\CO^{[n+1]}_\b\right\}\right) =
  \left(\left\{\CO^{[0]}_{\b}\right\}, \ldots,
    \left\{\CO^{[n]}_{\b}\right\}, \left\{\CO^{[n+1]}_{\b}\right\}
  \right).  }
$$
Note that $\CO^{[k+1]}_\a = \Fr{1}{(k+1)}t^\b \Delta\LL^{[k]}_{\b\a}$
for all $k=0,1,\ldots, n-1,n$ by assumption.  To prove above assertion
we need to establish that the relations \eqn\enika{ \sum_{j=0}^{k}
  \left(\CO^{[k-j+1]}, \CO^{[j]}_\b\right)=-Q\CO^{[k+1]}_{\b}, } for
all $k=0,1,\ldots, n$ - in particular $Q\CO^{[n+1]}_\a
=-\sum_{j=0}^{n}\left(\CO^{[n-j]},\CO^{[j]}_\a\right)$.  The above
assertion is a consequence of the identities \enik\ for $J_n$: we
apply $\Delta$ to the both hand-sides of the identities \enik\ for
$J_n$.  We have, for all $k=0,1,\ldots,n$ \eqn\enikb{
  (-1)^{|\a|}\sum_{j=0}^{k} \left(\CO^{[j]}_\a, \CO^{[k-j]}_\b\right)
  +\sum_{j=0}^{k-1}\left(\CO^{[k-j]},\Delta \LL^{[j]}_{\a\b}\right) =
  -Q\Delta \LL^{[k]}_{\a\b}, } where we used that $\Delta
{\CO}^{[\ell]}_\a =\Delta {\CO}^{[\ell+1]}=0$ for all
$\ell=0,1,\ldots, n$.  Multiplying both sides by $(-1)^\a t^\a$ and
sum over $\a$ to get
$$
\sum_{j=0}^{k} \left(t^\a \CO^{[j]}_\a, \CO^{[k-j]}_\b\right)
+\sum_{j=0}^{k-1}\left(\CO^{[k-j]},t^\a\Delta \LL^{[j]}_{\a\b}\right)
= -Qt^\a\Delta \LL^{[k]}_{\a\b},
$$
which gives
$$
\sum_{j=0}^{k} (j+1) \left(\CO^{[j+1]}, \CO^{[k-j]}_\b\right)
+\sum_{j=0}^{k-1}(j+1)\left(\CO^{[k-j]},\CO ^{[j]}_{\b}\right) =
-(k+1)Q\CO^{[k+1]}_{\b},
$$
for all $k=0,1,\ldots,n$. The left hand side of the above reduces to
$$
(k+1) \sum_{j=0}^{k} \left(\CO^{[k-j+1]}, \CO^{[j]}_\b\right)
$$
after a re-summation, which proves that the relations \enika\ are
satisfied.   

\end{itemize}

From the pair of inductive systems, it follows that every obstruction
vanishes and we have a solution $ \CO= \sum_{k=1}^\infty \CO^{[k]} \in
\left(\sC\otimes \Bbbk[[H^*]]\right)^0$ of the semi-classical master
equation such that $\CO^{[1]}= t^a \CO^{[0]}_\a=t^\a O_\a$, where the
cohomology classes of $\{O_\a\}$ from a basis of (total) cohomology of
the complex $(\sC, Q)$ and
$$
\CO^{[k+1]} :=\Fr{1}{(k+1)} t^\a \CO^{[k]}_\a,\qquad \CO^{[k+1]}_\a :=
\Fr{1}{(k+1)}t^\b \Delta\LL^{[k]}_{\b\a},
$$
for all $k=0,1,\ldots,\infty$. Note the above conditions are
equivalent to
$$
\left\{ \eqalign{ \CO^{[k]}_\a &= \Fr{\rd}{\rd t^\a}\CO^{[k+1]},\cr
    \Delta \LL^{[k]}_{\a\b} &=\Fr{\rd}{\rd t^\a}\CO^{[k+1]}_\b \equiv
    \Fr{\rd^2}{\rd t^\a\rd t^\b }\CO^{[k+2]},\cr }\right.
$$
for all $k\geq 0$. Thus $\CO_\a := \Fr{\rd}{\rd t^\a}\CO =
\sum_{k=0}^\infty \CO^{[k]}_\a$ and the system of equations \asuma\
limit is equivalent to the equation
$Q\CO_\a+\left(\CO,\CO_\a\right)=0$. Furthermore, $\CO_\a \in
\hbox{Ker }\Delta$ by construction.  Let $\LL_{\a\b} :=
\sum_{k=0}^\infty \LL^{[k]}_{\a\b}$ then we have $\CO_{\a\b} :=
\Fr{\rd^2}{\rd t^\a\rd t^\b}\CO= \Delta \LL_{\a\b}$ and the systems of
equations \asumb\ is equivalent to
$$
\CO_\a\cdot \CO_\b =\CA_{\a\b}{}^\g \CO_\g + Q\LL_{\a\b} +
(\CO,\LL_{\a\b}).
$$
Thus we have proved the proposition. \qed

\newsec{Further Developments}

The referee suggested to this author that Section 4 might be
simplified by using homological perturbation theory (HPT).  We note
that the paper \cite{HS} adopts HPT to reexamine the result of
Barannikov-Kontsevich \cite{BK}.  This author, however, doesn't have
any concrete ideas along the lines of this suggestion.

Consider a quantum algebra in Definition $4$ and call it anomaly-free
if every cohomology class of the associated classical complex has an
extension (or $\hbar$-correction) to a cohomology class of the quantum
complex.  The class of semi-classical quantum algebras considered in
this paper is a particular class of anomaly-free quantum algebras. A
generalization of this work to a anomaly-free quantum algebra shall be
studied in a sequel \cite{P2}.

\end{document}